\documentstyle[12pt,epsfig]{article}
\def\beq{\begin{equation}}
\def\eeq{\end{equation}}
\def\ber{\begin{eqnarray}}
\def\eer{\end{eqnarray}}

\def\a+{\alpha_+}
\def\beqn{\begin{eqnarray}}
\def\eeqn{\end{eqnarray}}

\begin{document}
\baselineskip=15.5pt
\pagestyle{plain}
\setcounter{page}{1}
\begin{titlepage}


\begin{center}

\vskip 4.7 cm

{\LARGE {\bf {Aspects of the free field description of\\
\vskip 0.8 cm
string theory on AdS$_3$}}} 

\vskip 2.1 cm

{\large Gast\'on Giribet and Carmen N\'u\~nez}

\vskip 1.0 cm

Instituto de Astronom\'{\i}a y F\'{\i}sica del Espacio

C.C. 67 - Suc. 28, 1428 Buenos Aires, Argentina

gaston, carmen@iafe.uba.ar

\vskip 2.1 cm

{\bf Abstract}

\end{center}
The near boundary limit of string theory in AdS$_3$ is analysed using
the Wakimoto free field representation of $SL(2,R)$. The theory is
considered as a direct product of the $SL(2,R)/U(1)$ coset
and a free boson. Correlation functions are constructed generalizing
to the non-compact case the integral representation of
conformal blocks introduced by Dotsenko in the compact $SU(2)$ CFT.
Sectors of the theory obtained by spectral
flow manifestly appear. 
The formalism naturally leads to consider scattering processes violating
winding number conservation. The consistency of the procedure 
is verified in the factorization limit.

\end{titlepage}

\newpage


\section{Introduction}

There are many motivations to study string theory in three dimensional 
Anti de Sitter 
spacetime. It was
realized more than one decade ago that an AdS$_3$ metric plus antisymmetric
tensor field provide an exact solution to the consistency conditions for
string propagation in non-trivial background fields. The corresponding
$\sigma$-model is a WZW model on the $SL(2,R)$ group manifold
(or on $SL(2,{\bf C})/SU(2) = H_3^+$ if one considers the Euclidean 
version).
More recently, the
AdS/CFT duality conjecture supplied an additional motivation. Since
both sides of the duality map, the three
dimensional string theory and the two dimensional CFT, are in principle 
completely solvable, this toy model raises the hope to 
explicitly work
out the details of the conjecture at the string level.

However, even though the theory has undergone a thorough examination over
the last years,
many important questions are still unanswered. In particular, it
is not yet clear what the spectrum of the theory is. 
The prescription to consider the principal continuous
series and the discrete representations (lowest and highest weight)
of $SL(2,R)$ (or its universal covering, ${\widetilde {SL(2,R)}}$) with
the
spin $j$ bounded by
unitarity leads to an unnatural limit on the level of excitation
of the
string states and to a partition function which is not modular invariant
(for a review and a complete list of references see
\cite{petro}), and it fails in the interacting theory \cite{gn}.

An interesting proposal was recently advanced by Maldacena and Ooguri
\cite{mo}. They realized that the $SL(2,R)$ WZW model has a spectral
flow symmetry which originates new admissible representations for the
string spectrum. Taking them into account the
problems mentioned above do not arise and it is possible to
consistently keep the
bound on the spin $j$  to avoid negative norm states in 
the free theory. 
This approach seems promising
and it would be very important to complete it by considering
interactions. In fact, a consistent string theory should provide
a mechanism to avoid ghosts at the interacting level, $i.e.$ non unitary
states should decouple in physical processes. But 
the computation of
correlation functions in this model presents several difficulties due to 
the non-compact nature of $H_3^+$, which
renders the proof of unitarity of the full theory highly non
trivial. Various attempts to include interactions have been developed in
recent years using different methods.
Certain correlators of the $SL(2,R)$ WZW model have been computed
by functional integration in 
\cite{gawedzki}. The bootstrap formalism was implemented in 
\cite{teschner} and 
two- and three-point functions for arbitrary spin $j$ were recently 
computed in \cite{satoh} using the path integral approach. 
The computation of higher point functions is 
important to completely
establish the consistency of the theory, but it gives rise to
 technical obstacles and complete expresions are not yet available.

Until more efficient calculational methods
emerge, the free field approach provides a useful tool to obtain 
some information.
It is suitable for describing
 processes in the near boundary
region of AdS$_3$
 (though results in \cite{satoh} suggest that it could apply to a larger
region). The  approach was used in \cite{gn} to study the
factorization
limit of $N$-point functions in the $H_3^+$ WZW model and determine the
unitarity of the theory. 
It is the purpose of this article to extend the free field formalism
to manifestly include the spectral flow symmetry. A direct extension 
of Dotsenko's 
method to compute the conformal blocks in the compact $SU(2)$
CFT \cite{dotsenko} to the non-compact $SL(2)$ (or $H_3^+$)
group manifold is found adequate to
deal with the spectral flow symmetry in vertex operators and scattering
processes 
and to describe interactions either conserving or violating
winding number 
conservation. In fact, the spectral flow parameter
$\omega$ is identified with the winding number of 
the string in AdS$_3$ 
and, as explained in references \cite{mo, gk}, 
it does not need to be conserved by interactions.

The general method carried out in the following sections to construct the
theory 
goes along the steps pursued in the proof of the no-ghost theorem
\cite{dpl, egp}. It begins
with the $H_3^+$ WZW model. Since the minus sign in the norm of some
states of the theory can be
traced to the $U(1)$ part of the current algebra, the states created by
the moments of this current are removed by considering the coset
$SL(2,R)/U(1)$. Finally, string theory in AdS$_3$ is recovered by
taking the tensor product of the coset with the state space of a
timelike free boson.

The paper is organized as follows. In Section 2 the free field 
description of SU(2) CFT
is reviewed by directly extending it to the non-compact case. The integral 
representation of the
conformal blocks and the mechanism to find the charge
 asymmetry conditions leading
to non-vanishing correlators is recalled.  
In Section 3 the quotient of $SL(2,R)$ by $U(1)$ is 
considered along the same path. The formalism naturally leads to find new
expressions for the vertex operators and new sets of charge asymmetry
conditions. This lays the ground to manifestly 
introduce the spectral flow symmetry into string theory on AdS$_3$ 
in Section 4, similarly as
what is done in the compact
case \cite{gepner}.  The scattering amplitudes for physical states are 
considered in Section
5 and their factorization properties are analysed in order to check the
consistency of the procedure. The vertex operators introduced in Section 3
are found useful to describe processes violating winding number
conservation. Finally the conclusions can be found in Section 6.

\section{Review of the free field representation of CFT}

In this section we review Dotsenko's construction of the free field
representation of  $SU(2)$ conformal field theory \cite{dotsenko} 
by extending it directly to the $SL(2)$ non-compact case.

The Wakimoto representation of $SL(2)$ current algebra \cite{waki} is
realized by three fields $\beta, \gamma, \phi$, with correlators
 given by
\beq
<\beta(z)\gamma(w)> = {1\over z-w} \quad ; \quad <\phi(z) \phi(w)> = -
{\rm ln}~(z-w)
\eeq

There are also $\bar z$ dependent antiholomorphic fields ($\bar\beta
(\bar z)$,
$\bar \gamma (\bar z)$, $\phi(\bar z)$). However we shall discuss the left
moving part of the theory only and assume that all the steps go through to
the right moving part as well, indicating the left-right matching
conditions where necessary.
 
The $SL(2)$ currents are represented as
\beqn
J^+(z) &=& \beta \nonumber\\
J^3(z) &=& -\beta\gamma - {\alpha_+\over 2}\partial\phi \nonumber\\
J^-(z) &=& \beta\gamma^2 + \alpha_+ \gamma\partial\phi + k\partial\gamma
\eeqn
where $\alpha_+ = \sqrt{2(k-2)}$ and $k$ is the level of the $SL(2)$
algebra.
They verify the following operator algebra
\begin{eqnarray}
J^{+}(z)J^{-}(w) &=&\frac k{(z-w)^2}-\frac 2{(z-w)}J^3(w)+{\rm RT}  \label{olgaz}
\\
J^3(z)J^{\pm}(w) &=&\pm\frac 1{(z-w)}J^{\pm}(w)+ {\rm RT}  \nonumber \\
J^3(z)J^3(w) &=&\frac{-k/2}{(z-w)^2}+ {\rm RT}  \nonumber
\end{eqnarray}
Expanding in Laurent series
\beq
J^a(z) = \sum_{n=-\infty}^\infty J_n^a ~z^{-n-1}
\eeq
the coefficients $J^a_n$ satisfy a Kac-Moody algebra given by
\beq
[J^a_n, J^b_m] = i\epsilon^{ab}_c J^c_{n+m} - {k\over 2}\eta^{ab} n
\delta_{n+m, 0} \label{ca}
\eeq
where the Cartan Killing metric is $\eta^{ab} = {\rm diag}(1,1,-1)$
and $\epsilon^{ab}_{c}$ is the Levi Civita antisymmetric tensor.

The Sugawara stress-energy tensor is
\beq
T_{SL(2)}(z) = \beta\partial\gamma - {1\over 2}\partial\phi\partial\phi -
{1\over \alpha_+} \partial^2\phi \label{tsl}
\eeq 
and it leads to the following central charge of the Virasoro algebra 
\beq
c= 3 + {12\over \alpha_+^2} = {3k\over k-2}. 
\eeq

The primary fields of the $SL(2)$ conformal theory $\Phi^j_m(z)$
satisfy the following OPE with the currents
\beqn
J^+(z) \Phi^j_m(w) &=& {(j-m)\over z-w} \Phi^j_{m+1}(w) + {\rm RT} \nonumber\\
J^3(z) \Phi^j_m(w) &=& {m\over z-w} \Phi^j_m(w) + {\rm RT} \nonumber\\
J^-(z) \Phi^j_m(w) &=& {(-j-m)\over z-w} \Phi^j_{m-1}(w) + {\rm RT}
\label{pf}
\eeqn

The corresponding vertex operators can be expressed as \cite{morozov}
\beq
\Phi^j_m(z) = \gamma^{j-m} e^{{2j\over \alpha_+}\phi}
\eeq
and their conformal dimensions are
\beq
\Delta(\Phi^j_m) = -{j(j+1)\over k-2} \label{cd}
\eeq

The next object of the free field realization is the screening
operator. It has to commute with all the currents, $i.e.$ it should
have no singular terms in the OPE with them. Up to a total derivative
this is satisfied by the operators \cite{bershad, morozov}
\beq
S_+(z) = \beta(z) e^{-{2\over \alpha_+}\phi} \quad ; \quad
S_-(z) = \beta ^{k-2} e^{-\a+\phi}
\label{scree}
\eeq
It can be checked that
\beqn
J^+(z) S_{\pm}(w) = {\rm RT} \quad &;& \quad J^3(z) S_{\pm}(w) = {\rm RT}
\nonumber\\
J^-(z) S_+(w) &=& (k-2)\partial_w \left ({e^{-{2\over \a+}\phi}\over
z-w}\right ) +
{\rm RT}\nonumber\\
J^-(z) S_-(w) &=& (k-2)\partial_w\left ({\beta^{k-3} e^{-\a+\phi}\over
z-w}\right ) + {\rm RT}
\eeqn
The total derivatives do not contribute if one integrates $S_\pm$ over
a closed contour. Then the screening operators
\beq
{\cal S}_\pm = \int_{\cal C} dz S_\pm(z) \label{scree}
\eeq
commute with the current algebra, they have zero conformal weight
 and can be used inside correlation functions without modifying their
conformal properties.

As shown by Dotsenko \cite{dotsenko}, to construct the integral
representation for the conformal blocks one needs a conjugate
operator for the fields $\Phi^j_m$ to avoid redundant
contour integrations which render the representation 
 incomplete. In order to find it, it is
important to construct the operator conjugate to the identity, which
determines the charge asymmetry conditions of the expectation values
in the radial-type quantization of the theory.
It has to commute with the currents and have conformal dimension
zero. It is found to be
\beq
\tilde {\cal I}_0 (z) = \beta^{k-1} e^{{2(1-k)\over \a+}\phi} \label{id0}
\eeq
Similarly as in the $SU(2)$ case one finds that there is no double
pole in the OPE  $J^-(z) \tilde {\cal I}_0(w)$ and that the residue of the
single
pole is
a spurious state which decouples in the conformal blocks for physical
states.

The conjugate identity operator requires that the charge asymmetry in
expectation values be
\beq
N_\beta - N_\gamma = k-1 \quad ; \quad \sum_i\alpha_i = {2-2k\over
\a+} \label{ca0}
\eeq
where $N_\beta (N_\gamma)$ refers to the number of $\beta (\gamma)$
fields in the correlator and $\alpha_i$ refers to the $``charge"$ of  
$\phi(z_i)$. Strong remarks against attributing the charge asymmetry
to the presence of the background charge operator in the expectation
values are given by Dotsenko \cite{dotsenko}.

One can now construct the conjugate representation for the highest
weight operators which turns out to be
\beq
\tilde \Phi^j_j(z) = \beta^{2j+k-1} e^{-{2(j-1+k)\over \a+}\phi}
\eeq
It can be checked that it satisfies the relations (\ref{pf}) corresponding
to a highest weight field ($i.e.,~ j=m$) and that its conformal dimension
is (\ref{cd}). Furthermore, it can be shown that the two-point
functions $<\tilde\Phi_j^j \Phi^j_{-j}>$ do not require screening
operators to satisfy the charge asymmetry conditions (\ref{ca0}).

The naive prescription to
compute the conformal blocks of the four-point functions, a
straightforward generalization of
the compact case, is
\beq
 <\Phi_{m_1}^{j_1}(z_1) \Phi_{m_2}^{j_2}(z_2) \Phi_{m_3}^{j_3} (z_3) 
\tilde\Phi_{j_4}^{j_4} (z_4)  \prod_i {\cal S}_+(u_i) \prod_j {\cal
S}_-(v_j)>
\label{pres}
\eeq
where the number of screening operators has to be chosen according to 
the charge asymmetry conditions (\ref{ca0}). Notice that it is possible
to satisfy them  using only one
type of
screening operators, namely ${\cal S}_+$. In the compact $SU(2)$ case it
seems convenient to use the conjugate representation operator in the
highest weight position for computation of conformal blocks and
correlation functions \cite{dotsenko} since the other operators of the
multiplet, $\tilde \Phi^j_m$, take more complicated expressions.

Conformal field theory based on ${\widehat {SL(2)}}_k$ has been studied
for fractional
levels of $k$ and spins in \cite{petko, andreev, rassmu}. 
Several technical difficulties arise from the
occurrence of fractional powers of $\beta, \gamma$ fields.
For applications to string theory in AdS$_3$ one needs to consider 
real values of the level $k$ satisfying $3 < c={3k\over k-2}\le 26$
(depending on the internal space). The spin $j$ is determined by the
mass shell and unitarity conditions. 
Let us briefly review this theory.

AdS$_3$ is the universal covering of the $SL(2,R)$ group
manifold (${\widetilde {SL(2,R)}}$). The sigma model describing string
propagation
in this
background plus an antisymmetric tensor field is a WZW model.
A well defined path integral formulation of the theory requires to
consider an Euclidean AdS$_3$ target space which is the $SL(2,{\bf
C})/SU(2) = H_3^+$ group manifold.
Using the Gauss parametrization,
the WZW model can be written as

\begin{equation}
S=k\int d^2z [\partial \phi \bar \partial \phi +\bar \partial \gamma
\partial 
\bar \gamma e^{2\phi }]  \label{s1}
\end{equation}
which describes strings propagating in three dimensional
Anti-de Sitter space with curvature $-\frac 2k$, Euclidean metric 
\begin{equation}
ds^2=kd\phi ^2+ke^{2\phi }d\gamma d\bar \gamma  \label{gik}
\end{equation}
and background antisymmetric field 
\begin{equation}
B=ke^{2\phi }d\gamma \wedge d\bar \gamma  \label{bik}
\end{equation}

The boundary of $AdS_3$ is located at $\phi \rightarrow \infty $. Near this
region quantum effects can be treated perturbatively, the exponent in the
last term in (\ref{s1}) is renormalized and a linear dilaton in $\phi $ is
generated. Adding auxiliary fields $(\beta ,\bar \beta )$ and rescaling, the
action becomes \cite{dfk} 
\begin{equation}
S=\frac 1{4\pi }\int d^2z [\partial \phi \bar \partial \phi -\frac
2{\alpha
_{+}}R\phi +\beta \bar \partial \gamma +\bar \beta \partial \bar \gamma
-\beta \bar \beta e^{-\frac 2{\alpha _{+}}\phi }].  \label{s2}
\end{equation}
 This description of the theory can be trusted for large
values of $\phi $. Note that the last term in (\ref{s2}) 
 is one of the screening operators (\ref{scree}). It is known
from the free field representation of the minimal models \cite{dotfat}
that the original Feigin-Fuchs prescription with contour integrals of
screening operators is equivalent to the one with the screening charges in
the action. It is usually assumed that the same equivalence holds in this
model \cite{dfk, becker}.

The string states must be in unitary representations of $SL(2, R)$
and satisfy the Virasoro constraints, $L_m\left| \Psi
\right\rangle =0$, $m>0$ and $L_o\left| \Psi \right\rangle =\left| \Psi
\right\rangle $. The last one implies 
\begin{equation}
-\frac{j(j+1)}{(k-2)}+L=1  \label{massshell}
\end{equation}
at excited level $L$. Notice that this expression is invariant under $%
j\rightarrow -j-1$.

Taking into account that the Casimir plays the role of mass squared
operator, the mass spectrum of the theory is 
\begin{equation}
M^2=\frac{(L-1)}2\alpha _{+}^2  \label{masita}
\end{equation}
Therefore, the ground state of the bosonic theory is a tachyon, the first
excited level contains massless states and there is an infinite tower of
massive states. If there is an internal compact space ${\cal N}$%
, eq. (\ref{massshell}) becomes 
\begin{equation}
-{\frac{j(j+1)}{{k-2}}} + L + h = 1  \label{msip}
\end{equation}
where $h$ is the contribution of the internal part.

Unlike string theory in Minkowski spacetime, the Virasoro constraints do
not decouple all the negative norm states. Since the physical spectrum of
string theory is expected to be unitary,
the admissible ${\widetilde {SL(2,R)}}$ 
representations are restricted.
Only the following unitary series at the base are relevant (see
\cite{mo}):

$i)$ Principal discrete highest weight representations:

\beq
{\cal D}_j^- = \{|j,m>; m = j, j-1, j-2, ...\}
\eeq
where $J_0^+|j,j>=0$.

$ii)$ Principal discrete lowest weight representations:

\beq
{\cal D}_j^+ = \{|j,m>; m = -j, -j-1, -j-2, ...\}
\eeq
where $J_0^-|j,-j>=0$. 

Unitarity requires $j\in R$ and
$-1/2 < j < {k-2\over 2}$ for both discrete series.

$iii)$ Principal continuous representation

\beq
{\cal C}_j = \{ |j,m>; j = -{1\over 2}+i\lambda;  \lambda, m \in R\}
\eeq
The full representation space is generated by acting on the states in
these series with $J_{n}^{\pm,3}$, $n<0$, and the corresponding
representations are denoted by ${\hat{\cal D}_j^\pm}, \hat{\cal C}_j$.

Correspondingly the conformal blocks that are relevant for string theory 
involve arbitrary (in general
complex or not positive integer) powers
of the ghost fields, and it is not obvious how to deal with them in
explicit calculations. 

 However other conceptual physical
problems are faced when one tries to identify this model with a string
theory. On the one hand, the bound on the spin of the discrete
representations implies an unnatural bound
on the
level of excitation of the string spectrum. Moreover, modular
transformations of the partition function of the theory defined in this
way reintroduce
states that are eliminated by the bound (see reference \cite{petro} for a 
complete discussion of these issues). Finally, interactions reintroduce,
in the intermediate channels, the negative norm states that were
eliminated by the unitarity bound \cite{gn}.

A natural solution to these problems was
recently proposed by Maldacena and Ooguri
\cite{mo}. They noticed an extra symmetry
in this theory which
allows to consider other representations of ${\widetilde {SL(2,R)}}$ in
addition to those mentioned above.
In order to extend the formalism reviewed in this section to that 
case, it
is convenient to first discuss the $SL(2,R)/U(1)$ coset theory.

\section{Extension to the $SL(2)/U(1)$ coset CFT}

The minus sign appearing in the norm of some states 
in representations of $SL(2, R)$ is associated with the $U(1)$ part of the
current algebra. Therefore a unitary module can be obtained by removing
all the states created by the moments of $J^3$. This procedure defines a
module for the coset
$SL(2,R)/U(1)$ and provides the basis for the proof of the no-ghost
theorem for string theory in AdS$_3$ \cite{dpl, egp}. In fact,  the
theorem is proved by first showing that all the
solutions to the Virasoro constraints on physical states can be
expressed as states in the $SL(2)/U(1)$ coset.
 Thus,  string theory in AdS$_3$ has a better chance
of being consistent if it is based on the coset model. 
Moreover, it was noticed in \cite{mo} 
that string theory on 
$AdS_3 \times {\cal N}$ is closely related to string theory on
${SL(2,R)\over U(1)} \times (time) \times {\cal N}$, the difference
lying in the conditions to be satisfied by the zero modes.
 Therefore it seems important to extend the formalism reviewed in the
previous section to the coset theory.

The $SL(2)/U(1)$ WZW theory describes string propagation in the
background of the two-dimensional black hole \cite{witten}. The
spectrum of this theory was discussed in \cite{dvv, dn} and certain
correlation functions where computed in \cite{becker}. Here
we follow a slightly different approach, based on Dotsenko's
formulation of the $SU(2)$ case, which will prove to be useful 
to manifestly include the spectral flow symmetry into 
string theory in AdS$_3$.

The procedure to gauge the $U(1)$ subgroup was  introduced in
reference \cite{dvv}. It amounts to adding a new free scalar field
$X$ and a ($b,c$) fermionic ghost system with propagators
\beq
<X(z) X(w)> = -{\rm ln}(z-w) \quad ; \quad <c(z) b(w)> = {1\over z-w}
\eeq

We are interested in the Euclidean theory, in which the boson $X$ is
compact with radius $R=\sqrt {k\over 2}$. 
The nilpotent BRST charge of this symmetry is
\beq
Q^{U(1)} = \int_{C_0} c (J^3 - i\sqrt {k\over 2}\partial X)
\label{brs}
\eeq
and the
stress-energy tensor of the gauged theory is
\beq
T_{SL(2)/U(1)} = T_{SL(2)} - {1\over 2}\partial X \partial X -
b\partial c \label{tcoset}
\eeq
where $T_{SL(2)}$ is given in (\ref{tsl}).

The primary fields of the coset theory should be
invariant under $Q^{U(1)}$. They are given by \cite{becker}
\beq
\Psi^j_m(z) = 
\gamma^{j-m} e^{{2j\over \alpha_+}\phi} e^{i\sqrt{2\over k}m X}
\label{pfc}
\eeq
and their conformal weight is
\beq
\Delta = -{j (j+1)\over k-2} + {m^2\over k}
\eeq

A comment on the antiholomorphic dependence of these fields is in order.
In the Euclidean black hole theory, the $J_0^3, \bar J_0^3$ eigenvalues
$m, \bar m$ lie on the lattice
\beq
m = {1\over 2} (p + \omega k) \quad ; \quad \bar m = -{1\over 2}
(p-\omega k)
\eeq
where $p$ (the discrete momentum of the string along the angular
direction) and $\omega$ (the winding number) are
integers.  The sum is $m+\bar m = \omega k$ and 
the difference $m-\bar m$ is an integer. This is to be contrasted
with the $\widetilde {SL(2,R)}$ case where $m+\bar m$ is not quantized. In
effect,
$m+\bar m$ is the spacetime energy of the string in AdS$_3$ and it may
take either discrete ($\hat{\cal D}_j^\pm$) or continuous ($\hat{\cal
C}_j$) values. 

 In order to construct the conformal blocks 
in the coset theory following the same steps
 as in the previous section, a secreening operator is needed.
It is evident from the expression (\ref{tcoset}) for the stress-energy
tensor that the screening operators in the coset theory are the same as in
the $SL(2)$ case, namely they are given by equations (\ref{scree}).

Next, the operator conjugate to the identity has to be found. Two
additional operators to that of the $SL(2)$ theory, $\tilde {\cal I}_0$ in
equation (\ref{id0}), exist (they
were introduced in reference \cite{mo}), namely
\beq
\tilde{\cal I}_+ = \gamma^{-k} e^{-{k\over \alpha_+}\phi}
e^{i\sqrt{k\over 2}X} \quad ; \quad \tilde{\cal I}_- = e^{-{k\over
\alpha_+}\phi} e^{-i\sqrt{k\over 2}X}
\eeq 

It is easy to check that they share the properties of $\tilde {\cal I}_0$,
$i.e.$
they commute with the currents and have vanishing conformal weight.
(Actually  
$J^+(z) \tilde {\cal I}_+(w)$ and $J^-(z) \tilde {\cal  I}_-(w)$ have a
non-vanishing
single pole, but a similar argument as the one made
 for $J^-(z) \tilde {\cal I}_0(w)$ applies, namely the residues are
spurious states which decouple in the conformal blocks).

Correspondingly two new sets of charge asymmetry conditions arise
\beqn
N_{\beta} - N_{\gamma} =  k \nonumber\\
\sum_i \alpha_i = - {k\over \alpha_+} \nonumber\\
\sqrt{2\over k} \sum_i \xi_i = \sqrt{k\over 2} \label{carga+}
\eeqn
and
\beqn
N_{\beta} - N_{\gamma} = 0  \nonumber\\
\sum_i \alpha_i = - {k\over \alpha_+} \nonumber\\
\sqrt{2\over k} \sum_i \xi_i = - \sqrt{k\over 2} \label{carga-}
\eeqn
Note that $\tilde {\cal I}_0$ given by (\ref{id0}) is also a good
conjugate
identity for the coset
theory, therefore equations (\ref{carga+}) and (\ref{carga-}) should
be completed in
this case as
\beqn
N_{\beta} - N_{\gamma} = k-1 \nonumber \\
\sum_i \alpha_i = {2-2k \over \alpha_+} \nonumber \\
\sum_i \xi_i = 0 \label{carga0}
\eeqn
$\xi_i$ denotes the $``charge"$ of the field $X(z_i)$.

Following the procedure outlined in the previous section to find the
integral representation of conformal blocks one needs the conjugate
representation of the highest weight fields. It is easy to show that the
following operators have the correct properties
\beq
\tilde \Psi^{j (0)}_j = 
\beta^{2j+k-1} e^{-{2(j-1+k)\over
\alpha_+}\phi}e^{i\sqrt {2\over k}j X}
\label{conj0}
\eeq
and 
\beq
\tilde \Psi^{j (-)}_j = \beta^{2j} e^{-{(2j+k)\over \a+}\phi}
e^{i\sqrt{2\over k}(j-{k\over 2}) X}
\label{conj-}
\eeq

One can check that the two-point functions 
$<\tilde\Psi^{j (0)}_j
\Psi^j_{-j}>_0$ and 
$<\tilde\Psi^{j (-)}_j
\Psi^j_{-j}>_-$  
do not require screening operators in order to satisfy equations
(\ref{carga0}) and (\ref{carga-}) respectively. Correspondingly, the
indices $(0)$ and
$(-)$ refer to the charge asymmetry conditions obtained from the conjugate
identities $\tilde {\cal I}_0$ and $\tilde {\cal I}_-$. 
Other conjugate operators in the multiplet $\Psi^j_m$ can be found by
acting with $J^-$ on the highest weight conjugate operator.
This construction mimics
the radial quantization in which the operator $\Psi$ creates one vacuum of
the Fock space, and $\tilde \Psi$ creates another vacuum, a conjugate
one (see reference \cite{dotsenko}). 

Therefore the $N$-point function in the coset theory takes the form
\beq
{\cal A}_N^{0,\pm} = <\prod_{i=1}^{N-1}\Psi^{j_i}_{m_i} (z_i) \tilde
\Psi_{j_N}^{j_N 
(0), (\pm)}(z_N) 
\prod_n {\cal S}_+(u_n)\prod_m {\cal S}_-(v_m)>_{0,\pm}
\label{corr}
\eeq
where the number of screening operators should satisfy the charge
asymmetry conditions
(\ref{carga0}), (\ref{carga+}) or
(\ref{carga-}), the conjugate highest weight operators are defined
accordingly and the corresponding amplitudes are denoted by 
${\cal A}_N^0$, ${\cal A}_N^+$ and ${\cal A}_N^-$,
respectively. It is easy to see that the conjugate operator 
in the sense of $\tilde {\cal I}_+$
does not have such a simple form as $\tilde\Psi^{j (0)}_j$ or $\tilde
\Psi^{j (-)}_j$, and thus 
the corresponding correlator ${\cal A}_N^+$ above should be taken as a
formal expression.

Notice that the conjugate operators (\ref{conj0}) and (\ref{conj-}) create
highest weight tachyons and can be used as vertex operators for such
states in the intermediate positions $z_2, ..., z_{N-1}$, $i.e.$ one can
insert up to $N-2$ conjugate operators of any kind in correlation
functions, as long as the in- and out- vacuum states are  
consistent with the scalar product between a direct and a conjugate
representation of the Fock space.

 At this point it is convenient to recall previous literature.
The connection between the free field description of the two dimensional
black hole and Liouville
theory plus $c=1$ matter was discussed in \cite{ms, bersha}. It was
shown that, as follows from naive counting of degrees of freedom, the
$\beta, \gamma$ system can be decoupled and all the results of this
section can be
phrased in terms of the $\phi, X$ fields, which describe a $c=1$ system
($X$) coupled to the Liouville mode ($\phi$). This implies that it might
be
possible to ignore the problems posed by non-integer powers of the ghost
fields and that analytic continuation in the fashion of \cite{dfk} 
can be used safely to compute the correlators (\ref{corr}).

Correlation functions for string theory in the black hole background were
computed in reference \cite{becker} by functional integration.
The equivalent of the
charge asymmetry conditions (\ref{carga0}) arise in that formalism from
the integral of the zero modes of the fields and the background charge in 
the action functional.
  In fact, recall
that $\Psi^{-j-1}_{-j}$ was used in \cite{becker} as the conjugate
highest weight operator, $\tilde \Psi_j^j$, and the following
charge asymmetry conditions were used: $N_\beta - N_\gamma = 1$;
$\sum_i\alpha_i =
-1$; $\sum_i m_i = 0$. 

\section{Spectral flow and vertex operators of string theory in AdS$_3$}

Let us now turn to string theory in AdS$_3$.
As noticed by Maldacena and Ooguri \cite{mo}, the algebra (\ref{ca})
has a
symmetry given by 
\begin{eqnarray}
J_n^3 &\rightarrow &\tilde J_n^3=J_n^3-\frac k2\omega \delta _{n,0} 
\nonumber \\
J_n^{\pm } &\rightarrow &\tilde J_n^{\pm }=J_{n\pm \omega }^{\pm }
\label{spectralflow}
\end{eqnarray}
where $\omega \in Z$. Consequently, the Laurent coefficients of the
stress-energy tensor
transform as
\begin{equation}
L_n\rightarrow \tilde L_n=L_n+\omega J_n^3-\frac k4\omega ^2\delta _{n,0}
\label{sf}
\end{equation}

This is the well known spectral flow symmetry \cite{schwim}. Classically
the parameter $\omega$ represents the winding number of the string
around the  center of AdS$_3$. Quantum mechanically one can define
asymptotic states consisting of
long strings whose winding number could
in principle change in a scattering process.

The spectral flow generates new representations of the
$SL(2,R)$
algebra. Indeed an eigenstate $\left| \tilde j,\tilde m\right\rangle $
of the operators $\tilde L_0$ and $\tilde J^3$ with the following
eigenvalues 
\begin{eqnarray*}
\tilde L_0\left| \tilde j,\tilde m\right\rangle &=&-\frac{\tilde j(\tilde
j+1)}{(k-2)}\left| \tilde j,\tilde m\right\rangle \\
\tilde J^3\left| \tilde j,\tilde m\right\rangle &=&\tilde m\left| \tilde
j,\tilde m\right\rangle
\end{eqnarray*}
is also an eigenstate of $L_0$ and $J^3$ with
eigenvalues given by 
\begin{eqnarray}
L_0\left|\tilde j, \tilde m\right\rangle &=&\left( -\frac{\tilde j(\tilde
j+1)}{(k-2)}%
-\omega \tilde m-\frac k4\omega ^2\right) \left| \tilde j,\tilde
m\right\rangle  \nonumber \\
J^3\left| \tilde j,\tilde m\right\rangle &=&\left( \tilde m+\frac k2\omega
\right) \left| \tilde j,\tilde m\right\rangle  \label{estas}
\end{eqnarray}

The Hilbert space of string theory in AdS$_3$ can be consequently extended
$\mathcal{H}\rightarrow \mathcal{H}_\omega 
$ in order to include the states $\left| \tilde j,\tilde
m, \omega\right\rangle $ obtained by spectral flow, which satisfy the
following on-shell condition 

\begin{equation}
(L_0-1)\left| \tilde j,\tilde m, \omega\right\rangle =\left( -\frac{\tilde
j(\tilde
j+1)}{(k-2)}-\omega \tilde m-\frac k4\omega ^2+L-1\right) \left| \tilde
j,\tilde m, \omega\right\rangle =0  \label{universall}
\end{equation}
(compare to equation (\ref{massshell}) and note that it is now possible
to consider bounds on the spin $\tilde j$ without limiting the excitation
level $L$). 
The new representations are denoted by $\hat{\cal
D}_{\tilde j}^{\pm,\omega}$ 
and ${\hat{\cal C}^\omega_{\tilde j}}$
and they consist of the spectral flow of the discrete (highest and lowest
weight) and continuous series respectively. These representations also
contain negative norm states, but Maldacena and Ooguri \cite {mo} have
shown that restricting the spin $\tilde j$ to $\tilde j<(k-2)/2$, the
Virasoro constraints remove
all the ghosts from the theory. Moreover, 
closure of the spectrum under
the spectral flow symmetry
implies that the upper unitarity bound on the spin $\tilde j$ of the
physical
states should be $\tilde j <{k-3 \over 2}$, $i.e.$ the bound is stronger
than 
required by the no-ghost theorem.

The spectrum of string theory consists then
 of a product of left and right representations 
$\hat{\cal C}^\omega_{\tilde j, L}\times \hat{\cal C}^\omega_{\tilde j, 
R}$ and 
$\hat{\cal D}_{\tilde j, L}^{\pm,\omega} \times \hat{\cal D}_{\tilde j,
R}^{\pm,\omega}$ 
with the same amount of
spectral flow and the same spin $\tilde j$ on the holomorphic and
antiholomorphic parts and with $-1/2<\tilde j<(k-3)/2$. The partition
function containing the spectral flow of the discrete representations with
this bound on the spin $\tilde j$ was shown to be modular invariant in
\cite{mo}. Moreover, the partition function for thermal AdS$_3$
backgrounds was also found to be modular invariant and consistent with
this spectrum in \cite{mos}. From now on we drop the tilde on $\tilde j, 
\tilde m$.

The spectral flow symmetry has
been extensively studied in the context of $N=2$ superconformal field
theories. Let us briefly review this case.
The $N=2$ superconformal algebra contains in addition to the Virasoro
generators $L_n$, two fermionic superpartners $G^\pm_n$ and a $U(1)$
current with Laurent coefficients $J_n$. The isomorphism of the algebras
generated by ($L_n, G^\pm_n, J_n$) 
and by the flowed ($\tilde L_n, \tilde G^\pm_n, \tilde J_n$) can be
interpreted in terms of the product of some quotient theory whose 
central charge is $c-1$ and a free scalar field which bosonizes the
$U(1)$ current. Indeed the $N=2$ generators
decompose into two mutually commuting sectors, one of which 
can be expressed in terms of the
parafermions defined by Zamolodchikov and Fateev \cite{zf} and the
other one  
contains a free boson. This observation led to establish
 the relation between the $N=2$ discrete series and the
representations of $SU(2)$ current algebra.
The generalization for $c>3$ was performed in
reference \cite{dpl} by considering the non-compact group $SL(2,R)$
and the corresponding parafermions introduced in reference \cite{lyk2}. 

In order to  implement this construction in string theory in AdS$_3$ it
seems natural to
consider the coset $SL(2,R)/U(1)$ (having central charge $c-1$) times a
free timelike scalar field $Y(z)$ which bosonizes 
the $J^3$ current as
\beq
J^3(z) = -i \sqrt{k\over 2} \partial Y(z)
\eeq
and has propagator $<Y(z) Y(w)> = {\rm ln} (z-w)$. 
However, instead of the parafermions,
one can use the Wakimoto representation introduced in the previous
section to describe the coset theory. In this
representation the energy-momentum tensor of the full theory takes the
form 
\begin{equation}
T =\beta \partial \gamma -\frac
12(\partial
\phi )^2-\frac 1{\alpha _{+}}\partial ^2\phi -\frac 12(\partial
X)^2-b\partial c+\frac 12(\partial Y)^2  \label{soten}
\end{equation}

Now a primary field with $J^3$ charge $m$ may be written as 
\beq
V^j_m = \Psi^j_m 
e^{i\sqrt{2\over k}mY(z)}.
\eeq
where  $\Psi^j_m$ is a $J^3$ neutral primary field in the coset theory
with conformal weight
\beq
\Delta(\Psi^j_m) = -{j(j+1)\over k-2} + {m^2\over k}.
\eeq

In terms of Wakimoto free fields  
it is  possible to write the corresponding 
vertex operators in the non-compact ${\widetilde {SL(2)}}$ case as (see
eq. (\ref{pfc})) 
\beq
V_m^j = \gamma^{j-m} e^{{2j\over \alpha_+}\phi} e^{i\sqrt{2\over k}m X}
e^{i\sqrt{2\over k}m Y(z)}.
\eeq

Now, taking into account the spectral flow,
for every field $V_m^j$ in the sector $\omega=0$ one can write a
field in the sector twisted by $\omega$ as
\beq
V^\omega_{j,m} = \gamma^{j-m} e^{{2j\over \alpha_+}\phi} e^{i\sqrt{2\over k}m X}
 e^{i\sqrt{2\over k} (m +\omega k/2)Y(z)}
 \label{mov}
\eeq
It has the following conformal weight 
\beq
\Delta(V^\omega_{j,m}) = -{j(j+1)\over k-2} - m\omega -
{k\omega^2\over 4}
\eeq
and therefore it has all the properties
to be considered the tachyon vertex operator in 
the free field representation of string theory in $AdS_3$. 

The general method proposed to construct the theory is then to begin with
the
local operators that create states of ${\widetilde {SL(2,R)}}$ and remove
the dependence
on the boson $X$. 
Once one has constructed the unitary modules for the coset, one can
combine them with the state space of a free boson $Y$ to build in unitary
representations of the full string theory on AdS$_3$.
Consequently the vertex operators are a direct product of an operator in
the
$SL(2)/U(1)$ coset theory  and an operator in the free field sector
representing the time direction (note the plus sign in the propagator
$<Y(z)Y(w)>$). 

There seem to be redundant degrees of
freedom in this representation. The situation is similar to the 
description of the two dimensional
black hole in terms of Wakimoto free fields plus a free boson. Therefore
it may be plausible that a simplified formulation exists also in this case
in
terms of only three fields \cite{ms, bersha}. An equivalent
expression for the vertex operators in terms of three free fields has been
recently introduced in
reference \cite{ponjas} in the discrete light-cone parametrization,
although a different interpretation is offered. However, both approaches
can be shown to be related upon using the constraint that $Q^{U(1)}$
annihilates the physical states of the coset theory. 

Now, in order to
complete the formulation of the theory, a prescription to compute
correlation functions is needed.

 \section{Scattering amplitudes and factorization}

The scattering amplitudes of physical states 
are essential ingredients  to obtain the spectrum
and study the unitarity of the theory. Several references have 
dealt with the problem
of computing correlation functions in $H^+_3$
\cite{gawedzki, teschner, satoh} and much 
progress has been achieved in recent years.
But it is difficult to construct higher than 
three-point functions without making some 
approximations. It would be interesting to resolve the 
technical problems in the evaluation of physical correlators as well as to
prove unitarity of string theory in AdS$_3$
at the interacting level.
In this section we take a step in this direction by extending
the
construction of
the free field representation of correlation functions discussed in the
context of the $SL(2)/U(1)$ coset in Section 3, to string theory in
AdS$_3$.

The free field approach is a powerful tool to study the theory
near the boundary of spacetime, even though the explicit computation of
correlation functions also presents some
technical difficulties. 
In particular, several
 properties can be obtained from
certain limits of the scattering amplitudes. Information about
 the spectrum is obtained
in the limit in which the insertion points of a subset of vertex
 operators collide to one point.
In this region of integration one finds
\beq
lim_{z_1,...,z_{M-1}\rightarrow z_M} {\cal A}_N = \sum_{L=0}^\infty
{{<V_1V_2...V_M\tilde V_{J_L}>
<V_{J_L} V_{M+1}...V_N>}\over {\Delta(V_{J_L})-1}}
\eeq
where $\Delta(V_{J_L})$ is the conformal dimension of the vertex operator
 creating the 
intermediate state at excitation level $L$. The consistency of the theory
can thus be established by starting with unitary external states and
analysing the norm of the intermediate states.

In order to implement this factorization process, the simplest
starting point is the
 scattering amplitude of unitary external tachyons.
Extending the ideas developed in Sections 2 and 3 to the case of 
string theory in 
AdS$_3$,
the $N$-point functions should take the following form
\beq
{\cal A}_N^{0,\pm} = <\prod_{i=1}^{N-1}V^{j_i}_{m_i,\omega_i}(z_i) \tilde
V^{j_N (0),(\pm)} _{j_N,\omega_N}(z_N)
\prod_{n=1}^s {\cal S}_+(u_n)>_{0,\pm} \label{cf}
\eeq
where the vertex operators $V^j_{m,\omega}$ are given in equation
(\ref{mov}) and the
conjugate highest weight operators are now
\beq
\tilde V^{j (0)}_{j,\omega} = \beta^{2j+k-1} e^{-{2(j-1+k)\over
\alpha_+}\phi}e^{i\sqrt {2\over k}j X} e^{i\sqrt{2\over k}(j+{k\over
2}\omega)Y} \label{fcon0}
\eeq
and
\beq
\tilde V^{j (-)}_{j,\omega} = \beta^{2j} e^{-{(2j+k)\over
\alpha_+}\phi}e^{i\sqrt {2\over k}(j-{k\over 2}) X} e^{i\sqrt{2\over
k}(j+{k\over
2}\omega)Y} \label{fcon-}
\eeq

Non-vanishing correlators require that the number of screening 
operators satisfy equations (\ref{carga+}), (\ref{carga-}) or (\ref{carga0})
plus an additional charge conservation condition arising from exponentials 
of the field $Y(z)$, namely
\beq
\sum \Omega_i = \sum_i (m_i + {\omega_i k\over 2}) = 0 \label{win}
\eeq
where $\Omega_i$ denotes the $``charge"$ of the field $Y(z_i)$.

This is the energy conservation condition.
In fact, $m+\bar m$ is the total energy of the string in
AdS$_3$ which receives kinetic as well as  winding contributions.
A similar condition arises
for the right moving part, and recalling that $\omega = \bar\omega$ is
implied by periodicity of the closed string coordinates, the left-right
matching condition is $\sum_i (m_i - \bar m_i) = 0$, $i.e.$ the angular
momentum conservation.
Consequently, without loss of generality, we shall consider states with
the same left and right quantum numbers.

It is interesting to notice that it is possible to construct correlators
violating winding number conservation by, for instance, inserting
conjugate operators
$\tilde V^{j (-)}_{j, \omega}$ instead of direct ones into
${\cal A}_N^{(0)}$.
In fact, correlation
functions containing $K$ of
these conjugate operators lead to $\sum_i\omega_i=-K$ when
combining
the last of equations (\ref{carga0}) with (\ref{win}), whereas 
processes conserving winding number ($\sum_i\omega_i = 0$) are obtained
when inserting direct
vertex operators.  Recall that it is possible to consider correlators
containing up to $N-2$
conjugate operators of a different kind as that required for the
conjugate vacuum state, and thus the winding number conservation can be
violated by up to $N-2$ units.
A similar observation was made
in reference \cite{gk}, where it is argued that
in the supersymmetric case the $N$-point functions receive
contributions that violate winding number conservation up to $N-2$. Moreover,
unpublished work by V. Fateev, A. B. Zamolodchikov and Al. B. Zamolodchikov
is quoted, where the same property seems to hold in the bosonic 
$SL(2)/U(1)$ CFT. 

In the
remaining of this article we shall check the consistency of the formalism
introduced in this Section by analysing the factorization properties of
the correlators. The procedure is very similar to the one introduced in
reference \cite{gn} and we include it here for completeness and to stress
the differences with the previous construction.

Let us start  from ${\cal A}_N^{(0)}$, $i.e.$ the $N$-point function for 
tachyons conserving winding number,
\begin{eqnarray*}
{\cal A}_{m_1...m_N}^{(0)j_1...j_N} &=&\frac 1{Vol[SL(2,C)]}%
\displaystyle \int 
\prod_{i=1}^Nd^2z_i%
\mathop{\displaystyle \prod }
_{n=1}^s%
\displaystyle \int 
d^2w_n\left\langle \prod_{i=1}^{N-1}\gamma _{(z_i)}^{j_i-m_i}
\beta_{(z_N)}^{2j_N+k-1}\prod_{n=1}^s\beta_
{(w_n)}\right\rangle \times c.c. \label{inel} \\
&&\ \  \times \left\langle
\prod_{i=1}^{N-1}e^{\frac 2{\alpha _{+}}j_i\phi (z_i,\bar z_i)}
e^{-2{(j_N-1+k)\over\a+}\phi(z_N, \bar z_N)}\prod_{n=1}^se^{-%
\frac 2{\alpha _{+}}\phi (w_n,\bar w_n)}\right\rangle \times \\
&&\ \ 
\times \left\langle \prod_{i=1}^{N-1} e^{i\sqrt{2\over k}m_iX(z_i,\bar
z_i)}
e^{i\sqrt{2\over k}j_N X(z_N,\bar z_N)} \right\rangle
\times \nonumber \\
&&\ \ \times \left\langle \prod_{i=1}^{N-1} e^{i\sqrt{2\over k}
(m_i+{k\over 2}\omega_i) Y(z_i, \bar z_i)}e^{i\sqrt{2\over k}
(j_N+{k\over 2}\omega_N) Y(z_N, \bar z_N)}
\right\rangle
\end{eqnarray*}
Here $(z_i,\bar z_i)$ and $(w_n,\bar w_n)$ are the world-sheet coordinates
where the tachyonic and the screening vertex operators, respectively, are
inserted. The quantum numbers of the external states and the number of
screening operators $s$ have to satisfy equations
(\ref{carga0}) and (\ref{win}). Note that we have taken the direct
representation for the vertex operators 
in intermediate positions $z_2,..., z_{N-1}$. However the discussion below
applies equally well (with minor modifications) to cases where one
considers some conjugate intermediate vertices.
 
Using the free field propagators this amplitude becomes
\begin{eqnarray}
{\cal A}_{m_1...m_N}^{(0)j_1...j_N} &\sim &%
\displaystyle \int 
\mathop{\displaystyle \prod }
_{i=1}^Nd^2z_i%
\mathop{\displaystyle \prod }
_{r=1}^sd^2w_n ~C(z_i,w_n) ~\bar C(\bar z_i,\bar w_n) \times \nonumber\\
&& \ \ \times \prod_{i<j=1}^{N-1}\left|
z_i-z_j\right| ^{-{8j_ij_j\over\alpha _{+}^2}+{4\over k}m_im_j - {4\over k}
(m_i+{k\over 2}\omega_i)
(m_j+{k\over 2}\omega_j)} \nonumber \\ 
&& \ \ \times \prod_{i=1}^{N-1}|z_i-z_N|^
{{8j_i(j_N-1+k)\over \a+^2} +
{4\over k}
m_ij_N 
- {4\over k}(m_i+{k\over 2}\omega_i)
(j_N+{k\over 2}\omega_N)} \nonumber \\
&& \ \ \times 
\prod_{i=1}^{N-1}\prod_{n=1}^s\left (\left| z_i-w_n\right| 
^{8j_i/\alpha _{+}^2}
|z_N-w_n|^{-8(j_N-1+k)/\a+^2}\right )\times
\prod_{n<m}^{s}\left| w_n-w_m\right| ^{-8/\alpha _{+}^2} \nonumber\\
 \label{tyuyt}
\end{eqnarray}
where $C(z_i,w_n)$ $[\bar C(\bar z_i,\bar w_n)]$ stand for the
contribution of the $(\beta ,\gamma )$ $[(\bar \beta, \bar\gamma)]$
correlators (see eq. (\ref{losc}) below).

Next take the limit $z_2\rightarrow 
z_1$. The amplitude is expected to exhibit poles on the mass-shell states
with residues reproducing the product of 3-point functions times 
$(N-1)-$point functions.
In this particular process there are three equivalent
possibilities, namely $lim_{z_2\rightarrow z_1} {\cal A}_N^{(0)}
\rightarrow $

\beqn
i&)& \sum_L{{<V_1 V_2 \tilde
V^{j_L(0)}_{j_L,\omega_L}>_0<V^{j_L}_{-j_L, -\omega_L} V_3 ...
\tilde
V^{(0)}_N>_0} 
\over {\Delta (V_{-j_L, -\omega_L}^{j_L}) - 1}} \label{fac0}\\
ii&)& \sum_L{{<V_1 V_2 \tilde V_{j_L,\omega_L}^{j_L,(+)}>_+
<V_{-j_L,-\omega_L}^{j_L (+)} V_3
... \tilde 
V^{(0)}_N>_0} 
\over {\Delta (V_{-j_L,-\omega_L}^{j_L}) - 1}} \label{fac+}\\
iii&)& \sum_L {{<V_1 V_2 \tilde V_{j_L,\omega_L}^{j_L
(-)}>_-<V_{-j_L,-\omega_L}^{j_L} V_3
...\tilde V^{(0)}_N>_0} 
\over {\Delta (V_{-j_L,-\omega_L}^{j_L}) - 1}} \label{fac-}
\eeqn
where the subindices refer to the different charge asymmetry conditions, 
$i.e.$ the number
of screening operators whose insertion points are taken in the limit
$w_n\rightarrow z_1$ in order to produce non-vanishing $3-$point functions
 verifies equations (\ref{carga0}), 
(\ref{carga+}) or
(\ref{carga-}), respectively (and the corresponding conjugate operator
has to be considered), 
and the remaining screenings in the $(N-1)-$point functions verify
conditions (\ref{carga0}).
$\Delta(V^{j_L}_{j_L,\omega_L})$ refer to
the conformal dimensions of the intermediate states at excitation level
$L$.
Even though we have explicitly constructed the correlators using the
conjugate operator in the highest weight position, the quantum numbers of
the intermediate states can be general ($i.e.$ not necessarily $j=m$). 

To isolate the
singularities arising in the intermediate channels perform the change
of variables: $z_1-z_2=\varepsilon ,$ $z_1-v_n=\varepsilon y_n,$ $%
v_n-z_2=\varepsilon (1-y_n)$, where we have renamed as $v_n$ the insertion
points of the $s_1$ screening operators that are necessary to produce
non-vanishing $3-$point functions. In order to extract 
the explicit $\varepsilon$ dependence of the amplitude it is convenient to
write the contribution of the $(\beta,\gamma)$ system as
\begin{eqnarray}
C(z_i,v_n, w_m) &=&\left\langle \gamma ^{j_1-m_1}_{(z_1)}
\gamma ^{j_2-m_2}_{(z_2)}%
\mathop{\displaystyle \prod }
_{i=3}^{N-1}\gamma ^{j_i-m_i}_{(z_i)} \beta_{(z_N)}^{2j_N+k-1}
\mathop{\displaystyle \prod }
_{n=1}^{s_1}\beta_{(v_n)}%
\mathop{\displaystyle \prod }
_{m=1}^{s_2}\beta_{(w_m)}\right\rangle  \nonumber \\
\ &\sim&%
\mathop{\displaystyle \sum }
_{Perm(v_n)}%
\mathop{\displaystyle \sum }
_{r=0}^{s_1}\frac{(j_1-m_1)(j_1-m_1-1)...(j_1-m_1-r+1)}{%
(z_1-v_1)(z_1-v_2)...(z_1-v_r)}\times  \nonumber 
\end{eqnarray}
\begin{eqnarray}
&&~~~~~~~~~~~~~~~~~~~~~~~ 
\times \frac{(j_2-m_2)(j_2-m_2-1)..(j_2-m_2-s_1+r+1)}{%
(z_2-v_{r+1})...(z_2-v_{s_1})}\times  \nonumber \\
&& \
 \times \left\langle \gamma ^{j_1-m_1-r}_{(z_1)}
\gamma ^{j_2-m_2-s_1+r}_{(z_2)}%
\mathop{\displaystyle \prod }
_{i=3}^{N-1}\gamma ^{j_i-m_i}_{(z_i)} \beta_{(z_N)}^{2j_N+k-1}
\mathop{\displaystyle \prod }
_{m=1}^{s_2}\beta_{(w_m)}\right\rangle +  \nonumber \\
&&\ \ +%
\mathop{\displaystyle \sum }
_{Perm(v_n)}%
\mathop{\displaystyle \sum }
_{r=0}^{s_1-1}\frac{(j_1-m_1)(j_1-m_1-1)...(j_1-m_1-r+1)}{%
(z_1-v_1)(z_1-v_2)...(z_1-v_r)}\times  \nonumber \\
&&\ \ \times \frac{(j_2-m_2)(j_2-m_2-1)..(j_2-m_2-s_1+r+2)}{%
(z_2-v_{r+1})...(z_2-v_{s_1-1})}%
\mathop{\displaystyle \sum }
_{i=3}^N\frac{j_i-m_i}{z_i-v_{s_1}}\times  \nonumber \\
&&\ \ \times \left\langle \gamma ^{j_1-m_1-r}_{(z_1)}\gamma
^{j_2-m_2-s_1+r+1}_{(z_2)}%
\mathop{\displaystyle \prod }
_{l\neq i}^{}\gamma ^{j_l-m_l}_{(z_l)}\gamma ^{j_i-m_i-1}_{(z_i)}%
\beta_{(z_N)}^{2j_N+k-1}
\mathop{\displaystyle \prod }
_{m=1}^{s_2}\beta_{(w_m)}\right\rangle +...\nonumber\\ 
\label{losc}
\end{eqnarray}
and similarly for $\bar C(\bar z_i,\bar v_n,\bar
w_n)$.
 The sign $\sim$
stands for an irrelevant phase. The products $%
(j_1-m_1)(j_1-m_1-1)...(j_1-m_1-r+1)$ have to be understood as not
contributing for $r=0$ (similarly $(j_2-m_2)...(j_2-m_2-s_1+r+1)$ for
$r=s_1$). 
The dots stand for lower order contractions between the fields inserted
at $z_1$ and $z_2$ and the $s_1$ screening operators. Note that these
functions can be written as a power series in $\varepsilon $ after
performing the change of variables and extracting the leading $\varepsilon
^{-s_1}$ divergence. Note that these expressions are obtained by treating
the powers of the $\beta, \gamma$ fields as positive integers and
assuming that analytic continuation can be safely performed.

The amplitude becomes then formally in the limit
\begin{eqnarray}
{\cal A}_{m_1...m_N}^{(0)j_1...j_N} &\sim &%
\displaystyle \int 
d^2\varepsilon \left| \varepsilon \right|
^{2s_1-{1\over \a+^2}[8j_1j_2-8s_1(j_1+j_2)+4s_1(s_1-1)]-2
(m_1\omega_2+m_2\omega_1 +\omega_1\omega_2 {k\over 2})-2s_1}\times 
\nonumber \\
&&\times 
\displaystyle \int 
d^2z_1\prod_{i=3}^N%
\displaystyle \int 
d^2z_i%
\mathop{\displaystyle \prod }
_{n=1}^{s_2} \int d^2w_n%
\mathop{\displaystyle \prod }
_{r=1}^{s_1}%
\displaystyle \int 
d^2y_r\left| \Phi (\varepsilon,z_1,z_i,y_r,w_n)\Psi (z_i,w_n)
\right|^2\nonumber\\
\label{yy}
\end{eqnarray}

The first term in the exponent of $\left| \varepsilon \right| $ comes from
the change of variables in the insertion points of the $s_1$ screening
operators whereas the last term cancelling it arises in the $\beta-\gamma$
system.
The other terms in the exponent originate in the
contractions of the exponentials. The function $\Phi $ is a regular function
in the limit $\varepsilon \rightarrow 0$. It is convenient to write
separately the contribution to $\Phi$ from the exponentials ($E$) and from
the $\beta - \gamma$ system ($C$), $i.e.$ $\Phi = E\times |C|^2$, where 
$E(\varepsilon, z_1, z_i, y_r, w_n)$ is
\begin{eqnarray}
E =&&\prod_{r=1}^{s_1}
\ |y_r|^{8j_1/\alpha_+^2} |1-y_r|^{8j_2/\alpha_+^2}
\prod_{r<t} \ |y_r-y_t|^{-8/\alpha_+^2}  \nonumber \\
&&\prod_{i=3}^{N-1} |z_1 - z_i|^{-8j_1j_i/\alpha_+^2-2m_1\omega_i-2\omega_1m_i
-k\omega_1\omega_i} |z_1 - \varepsilon -
z_i|^{-8j_2 j_i/\alpha_+^2-2m_2\omega_i-2\omega_2m_i
-k\omega_2\omega_i}  \nonumber \\
&&\ \ \ \ ~~  \times
|z_1-z_N|^{8j_1(j_N-1+k)/\a+^2-2(m_1\omega_N+\omega_1j_N)-k\omega_1\omega_N}
\nonumber \\
&& \ \ \ ~~
\times 
|z_1-\varepsilon-z_N|^{8j_2(j_N-1+k)/\a+^2-2(m_2\omega_N+\omega_2j_N)
-k\omega_2\omega_N} \nonumber \\
&&\prod_{m=1}^{s_2} \ |z_1 - w_m| ^{8j_1/\alpha_+^2} \ |z_1 - \varepsilon -
w_m|^{8j_2/\alpha_+^2}  \nonumber \\
&&\prod_{i=3}^{N-1}\prod_{r=1}^{s_1} \ |z_i - z_1 + \varepsilon y_r
|^{8j_i/\alpha_+^2} \prod_{r=1}^{s_1}\prod_{m=1}^{s_2} \ |z_1 - \varepsilon
y_r - w_m|^{-8/\alpha_+^2} \nonumber\\
&&\prod_{r=1}^{s_1}|z_N - z_1 + \varepsilon y_r
|^{-8(j_N-1+k)/\alpha_+^2} \nonumber \\ 
\label{phi}
\end{eqnarray}
and 
\begin{eqnarray}
&&C(\varepsilon, z_1, z_i, y_r, w_n) \sim \mathop{\displaystyle \sum }
_{Perm(y_n)}%
\mathop{\displaystyle \sum }
_{r=0}^{s_1}\frac{(j_1-m_1)(j_1-m_1-1)...(j_1-m_1-r+1)}{y_1y_2...y_r}\times 
\nonumber \\
&&~~~~~~~~~~~~~~~~~~ \times
\frac{(j_2-m_2)(j_2-m_2-1)...(j_2-m_2-s_1+r+1)}{%
(1-y_{r+1})(1-y_{r+2})...(1-y_{s_1})}\times  \nonumber \\
&&
\sum_{Perm(w_m)}{\large \{}{\large [ }{-\frac{(j_1-m_1-r)}{z_1-w_1}} - 
\frac{(j_2-m_2-s_1+r)} {(z_1-\varepsilon -w_1)} {\large ]} 
<\prod_{i=3}^{N-1}         
\gamma^{j_i-m_i}_{(z_i)} \beta^{2j_N+k-1}_{(z_N)} 
\prod_{m=2}^{s_2}\beta_{(w_m)}> +  \nonumber \\
&&+ {\large [} \frac{(j_1-m_1-r)(j_1-m_1-r-1)}{(z_1 - w_1)(z_1-w_2)} +
\frac{%
(j_2-m_2-s_1+r)(j_2-m_2-s_1+r-1)}{(z_1-\varepsilon-w_1)(z_1-\varepsilon
-w_2)%
}  \nonumber \\
&&~~~~~~~~~~~~~
+ \frac{(j_1-m_1-r)(j_2-m_2-s_1+r)}{(z_1-w_1)(z_1-\varepsilon-w_2)}
+\frac{%
(j_1-m_1-r)(j_2-m_2-s_1+r)}{(z_1-w_2)(z_1-\varepsilon-w_1)} {\large ]} 
\nonumber \\
&&~~~~~~~~~~~~~~~~~~~~~<\prod_{i=3}^{N-1} \gamma^{j_i-m_i}(z_i)
\beta^{2j_N+k-1}(z_N)\prod_{m=3}^{s_2}\beta (w_m)> +... {\large \}} + ...
  \label{bg}
\end{eqnarray}
The dots inside the bracket in the last equation stand for terms involving
more contractions among the vertices at $z_1$ and $z_2$ and the vertex
operator at $z_N$ or the $s_2$
screenings at $w_m$, whereas the dots at the end stand for lower order
contractions between the colliding vertices ($V^{\omega_1}_{(j_1,m_1)}$
and $V^{\omega_2}_{(j_2,m_2)}$) and the $s_1$ screenings at $v_n$.

The function $\Psi $ in eq. (\ref{yy}) is independent of $\varepsilon $.

It is possible to Laurent expand $\Phi $ as 
\begin{equation}
\Phi =%
\mathop{\displaystyle \sum }
_{n,m,l,\bar l}\frac 1{n!m!}\varepsilon ^{n+l}\bar \varepsilon ^{m+\bar l%
}\partial ^n\bar \partial ^m\Phi _{l\bar l}{}_{\mid \varepsilon =\bar 
\varepsilon =0}  \nonumber
\end{equation}
where $\Phi _{l\bar l}$ denotes the contributions from terms in $%
C(z_i,v_n,w_m)$ and $\bar C(\bar z_i,\bar v_n, \bar w_m)$ where a number $l$
($\bar l$) of the $\beta$-fields $(\bar\beta)$ in the $s_1$ screenings are
not contracted with the $\gamma$-fields $(\bar\gamma)$ of the vertices at $%
z_1$ and $z_2$, but with the other vertices at $z_i$, $i=3,..., N-1$.

Inserting this expansion in (\ref{yy}) and performing the integral over $%
\varepsilon $, the result is ${\cal A}_{m_1...m_N}^{(0)j_1...j_N} \sim $
\begin{eqnarray}
&&\sim \mathop{\displaystyle \sum }
_{n,l}\frac{\Lambda ^{[-8j_1j_2+8s_1(j_1+j_2)-4s_1(s_1-1)]/\alpha
_{+}^2-2(m_1\omega_2+m_2\omega_1+\omega_1\omega_2k/2)+2n+2l+2}}
{-\frac 8{\alpha _{+}^2}j_1j_2+\frac 8{\alpha _{+}^2}%
s_1(j_1+j_2)-\frac 4{\alpha _{+}^2}s_1(s_1-1)-2(m_1\omega_2+m_2\omega_1+
\omega_1\omega_2{k\over 2})+2n+2l+2}  \nonumber \\
&&\ \times 
\displaystyle \int 
d^2z_1%
\mathop{\displaystyle \prod }
_{t=1}^{s_2}%
\displaystyle \int 
d^2w_t\prod_{i=3}^N%
\displaystyle \int 
d^2z_i\prod_{r=1}^{s_1}%
\displaystyle \int 
d^2y_r\frac 1{(n!)^2}\partial ^n\Phi _l{}_{\mid \varepsilon =0}\Psi
(z_i,w_t)\times c.c. \nonumber \\
 \label{jkjkjk}
\end{eqnarray}
where $\Lambda $ is an infrared cut-off, irrelevant on the poles.

Let us analyse the pole structure of this expression, namely 
\begin{equation}
-\frac 4{\alpha _{+}^2}j_1j_2+\frac 4{\alpha _{+}^2}s_1(j_1+j_2)-\frac 2{%
\alpha_{+}^2}s_1(s_1-1)-(m_1\omega_2+m_2\omega_1+
\omega_1\omega_2k/2)+1+n+l=0  \label{shouldg}
\end{equation}

This is precisely the mass shell condition for a 
highest weight state at 
level $L=n+l$
 with $ ~ j=j_1+j_2-s_1$,  $m= m_1 + m_2$ and
$\omega = \omega_1 + \omega_2$, $i.e.$ 
\begin{equation}
-\frac 2{\alpha _{+}^2}j(j+1) - m\omega - {k\over 4}\omega^2 +L =
1  \label{shouldk}
\end{equation}
if $j_1, m_1, \omega_1$ and $j_2,m_2,\omega_2$ 
are the quantum numbers of the external 
on-mass-shell tachyons (namely, $-\frac{2j_1(j_1+1)}{%
\alpha _{+}^2}-m_1\omega_1-{k\over 4}\omega_1^2=1$). 

 At this point it is
important to recall that the scattering amplitudes (and the charge
asymmetry conditions) are constructed 
including one conjugate highest weight field. In general, the conjugate
vertex operator
is a complicated expression, except for the highest
weight state and this is the reason why these particular correlators are
considered. Therefore, the consistency of the factorization procedure,
$i.e.$ the non trivial fact that the number of screening operators
contained in the original amplitude can be split in two parts giving rise
exactly to two non-vanishing correlators in the residues,
has been checked explicitly in the special case in which the intermediate
states
verify $j=m$. However, more general processes might be
considered ($i.e.$ not necessarily containing highest weight states).

Next, let us consider the residues.
At lowest order ($n=l=0$), the amplitude
${\cal A}_{m_1...m_N}^{(0)j_1...j_N}
(\varepsilon =0)$ reads
\begin{eqnarray}
\mathop{\displaystyle \prod }
_{r=1}^{s_1}%
\displaystyle \int 
d^2y_r%
&&\mathop{\displaystyle \prod }
_{r=1}^{s_1}\left| y_r\right| ^{8j_1/\alpha _{+}^2}\left| 1-y_r\right|
^{8j_2/\alpha _{+}^2}%
\mathop{\displaystyle \prod }
_{r<t}^{s_1}\left| y_r-y_t\right| ^{-8/\alpha _{+}^2}\times C^{\prime }(y_r)%
\bar C^{\prime }(\bar y_r)\times  \nonumber \\
&& \times 
\displaystyle \int 
d^2z_1%
\displaystyle \int 
\mathop{\displaystyle \prod }
_{i=3}^Nd^2z_i\prod_{n=1}^{s_2}%
\displaystyle \int 
d^2w_n\left| z_1-z_i\right| ^{-8(j_1+j_2-s_1)j_i/\alpha
_{+}^2-2(m_1+m_2)\omega_i-(\omega_1+\omega_2)(2m_i+k\omega_i)}\times 
\nonumber \\
&& \times \prod_{n=1}^{s_2}\left| z_1-w_n\right| ^{8(j_1+j_2-s_1)/\alpha
_{+}^2}%
\mathop{\displaystyle \prod }
_{3<i<k}^{N-1}\left| z_i-z_k\right| ^{-8j_ij_k/\alpha _{+}^2}%
\mathop{\displaystyle \prod }
_{i=3}^{N-1}%
\mathop{\displaystyle \prod }
_{n=1}^{s_2}\left| z_i-w_n\right| ^{8j_i/\alpha _{+}^2}\times  \nonumber
\\
&& \times
\mathop{\displaystyle \prod }_{i=1}^{N-1} 
|z_i-z_N|^{{{8j_i(j_N-1+k)\over \a+^2} +
{4\over k}
m_ij_N 
- {4\over k}(m_i+{k\over 2}\omega_i)
(j_N+{k\over 2}\omega_N)}} 
\prod_{n=1}^{s_2} |z_N-w_n|^{-8(j_N-1+k)/\a+^2} \nonumber \\
&& \times 
\mathop{\displaystyle \prod }
_{n<m}^{s_2}\left| w_n-w_m\right| ^{-8/\alpha _{+}^2}\times C^{\prime
\prime
}(z_1,z_i,w_n)\bar C^{\prime \prime }(\bar z_1,\bar z_i,\bar w_n)
\label{popz}
\end{eqnarray}
where 
\begin{eqnarray*}
C^{\prime }(y_r) &=&%
\mathop{\displaystyle \sum }
_{Perm(y_n)}%
\mathop{\displaystyle \sum }
_{r=0}^{s_1}\frac{(j_1-m_1)(j_1-m_1-1)...(j_1-m_1-r+1)}{y_1y_2...y_r}\times
\\
&&\ \times \frac{(j_2-m_2)(j_2-m_2-1)...(j_2-m_2-s_1+r+1)}{%
(1-y_{r+1})(1-y_{r+2})...(1-y_{s_1})}
\end{eqnarray*}
and clearly from eq. (\ref{bg}) evaluated at $\varepsilon = 0$, 
\[
C^{\prime \prime }(z_1,z_i,w_n)=\left\langle \gamma
^{j_1-m_1+j_2-m_2-s_1}_{(z_1)}%
\mathop{\displaystyle \prod }
_{i=3}^{N-1}\gamma ^{j_i-m_i}_{(z_i)} \beta_{(z_N)}^{2j_N+k-1}%
\mathop{\displaystyle \prod }
_{m=1}^{s_2}\beta_{(w_m)}\right\rangle 
\]

This can be easily interpreted as the product of a $3$-tachyon amplitude
(the first line in expression (\ref{popz})) 
\begin{equation}
\left\langle V^{\omega_1}_{j_1,m_1}(0)V^{\omega_2}_{j_2,m_2}(1)\tilde V%
^\omega_{j,m}(\infty )%
\mathop{\displaystyle \prod }
_{r=1}^{s_1}{\cal S}_+(y_r)\right\rangle  \label{sallaw}
\end{equation}
times a $(N-1)$-tachyon amplitude 
\begin{eqnarray}
&&\left\langle \gamma _{(z_1)}^{(j_1+j_2-s_1)-m_1-m_2}\gamma
_{(z_3)}^{j_3-m_3}...\beta _{(z_N)}^{2j_N+k-1}\prod_{n=1}^{s_2}\beta_
{(w_n)}\right\rangle \times  \label{wallas} \\
&&\times \left\langle \bar \gamma_{(\bar z_1)}^{(j_1+j_2-s_1)-
m_1-m_2}\bar 
\gamma _{(\bar z_3)}^{j_3- m_3}...\bar \beta _{(\bar z_N)}^{2j_N+k-1}%
\prod_{r=1}^{s_2}\bar \beta_{(\bar w_n)}\right\rangle \times  \nonumber \\
&&\times \left\langle e^{2(j_1+j_2-s_1)\phi (z_1,\bar z_1)/\alpha
_{+}}\prod_{i=3}^{N-1} e^{2j_i\phi (z_i,\bar z_i)/\alpha
_{+}} e^{-2(j_N-1+k)\phi(z_N,\bar z_N)/\a+}
\prod_{n=1}^{s_2}e^{-2\phi (w_n,\bar w_n)/\alpha _{+}}\right\rangle \times 
\nonumber \\
&&\times \left\langle e^{i\sqrt{2\over
k}(m_1+m_2)X(z_1,\bar
z_1)} \prod_{i=3}^{N-1}e^{i\sqrt{2\over k}m_iX(z_i,\bar z_i)}
e^{i\sqrt{2\over k}j_N X(z_N,\bar z_N)} \right\rangle \times \nonumber \\
&& \times \left\langle e^{i\sqrt{2\over
k}(m_1+m_2+(\omega_1+\omega_2){k\over
2})Y(z_1,\bar z_1)} \prod_{i=3}^{N-1}e^{i\sqrt{2\over k}(m_i+{k\over
2}\omega_i)Y(z_i,\bar z_i)} e^{i\sqrt{2\over k}(j_N+{k\over 2}\omega_N)}
\right\rangle
\end{eqnarray}

Therefore, the tachyon vertex operator can be reconstructed, namely
\begin{equation}
V^\omega_{(j,m)}(z,\bar z)
 ~ = ~ \gamma ^{j-m}(z)\bar \gamma ^{j-m}(\bar z)e^{\frac 2{\alpha
_{+}}j\phi (z,\bar z)} e^{i\sqrt{2\over k}m X(z,\bar z)}
e^{i\sqrt{2\over k}(m+ {k \over 2}\omega) Y(z,\bar z)} 
\label{tachyk}
\end{equation}
with $j=j_1+j_2-s_1$, $m=m_1+m_2$ and
$\omega=\omega_1+\omega_2$. 

The vertex operators creating states at higher excitation levels can be
obtained from the higher order terms in the Laurent expansion
(\ref{jkjkjk}) following the same steps implemented in this section.

\section{Conclusions and discussion}

The near boundary limit of string theory in AdS$_3$ has been considered
using the Wakimoto free field representation of ${\widetilde {SL(2, R)}}$.
The
theory was
taken as a direct product of the $SL(2,R)/U(1)$ coset and a timelike
free boson. The winding sectors obtained by the spectral flow
transformation appear naturally in the spectrum of the theory.
Correlation functions of physical states were constructed
extending to the non-compact case, Dotsenko's integral representation of
conformal blocks in the $SU(2)$ case. There are three sets of charge
asymmetry conditions
arising from the corresponding conjugate identity operators.
Conjugate vertex operators can be constructed with the help of these
conditions, and they can be used to
describe
scattering processes either conserving or violating winding number
(in the latter case by up to $N-2$ units). 
We have explicitly constructed these conjugate operators for the highest
weight states and indicated the procedure that should be followed to find
more general conjugate operators. The consistency of the formalism
was checked in the 
factorization
limit obtained when the insertion points of two external vertex operators 
coincide on the sphere. In this limit the amplitudes were shown to exhibit
poles on
the mass-shell states and residues reproducing the products of $3-$ and 
$(N-1)-$point functions of the external states with the intermediate
on-mass
shell states.

This formalism can be used to compute scattering amplitudes and to study
the unitarity of string
theory in AdS$_3$ at the
interacting level, similarly as was done in reference \cite {gn}. In
fact, starting from the scattering amplitudes of
unitary external states, one can analyse the quantum numbers
of the intermediate states and determine if they fall within the unitarity
bound. However the procedure has some limitations and more work is
necessary to achieve this goal, as well as to explicitly evaluate the
correlators. 

The radial-type quantization that we have considered
requires a conjugate field defining a conjugate vacuum state. The vertex
operator
for such field is in general a complicated expression, unless it is a
highest weight field. In order to find
unitary highest weight
tachyons in string theory one has to take into account an internal compact
space,
$i.e.$ string theory on AdS$_3 \times {\cal N}$ has to be considered, and 
this requires working out an explicit example. 

On the other hand, if one is interested in applications
to string theory, the asymptotic states naturally leading to the
definition of the S-matrix consist of long strings. The states
describing the long strings belong to the spectral flow of the continuous
representation \cite{mo} while the correlators discussed in this
article contain at least one highest weight field. It is usually assumed
that an analytical continuation can be performed and that both real or 
complex values of the spin $j$ can be treated on an equal footing. But an
explicit calculation of mixed fusion rules (involving states both in the
continuous and discrete representations) requires the
functional form of the conjugate operators for all representations,
including the continuous series. Moreover there is no clear 
physical interpretation of the role played by the
discrete representations in the scattering amplitudes of asymptotic string
states. The extension of the objects constructed in this article to
take into account states of the continuous representation is necessary for
a complete understanding of the theory.

Another aspect of the construction which requires more thought is the
absence of simple conjugate highest weight operators with
respect to
the charge asymmetry conditions (\ref{carga+}). It would be important to
have explicit expressions for these operators since it is
likely that they can be used to describe scattering amplitudes
violating
winding number conservation by a positive integer.

Explicit evaluation of $3-$point functions with
this formalism would be important in order to compare with other
approaches
\cite{teschner,
satoh}. Furthermore the
extension of the construction presented here to superstring theory
should be interesting.

 \bigskip
{\bf{Acknowledgements}}

We would like to thank J. Maldacena for useful discussions.
C.N. is grateful to the International Centre for Theoretical Physics for
hospitality during the period in which part of this work was elaborated.
This work was supported in part by grants from CONICET (PIP 0873) and
ANPCyT (PICT-0303403), Argentina.

\end{document}